\documentclass[10pt]{article}
\usepackage{makeidx,epsfig,lscape}
\usepackage[hyperindex]{hyperref}
\usepackage{color,colortbl}
\usepackage{fancyhdr}
\usepackage{xcolor,pict2e}
\usepackage{authblk}
\usepackage{amsmath,amssymb,amsthm,amsfonts}
\usepackage{mathrsfs,bm}
\usepackage{dsfont}
\usepackage{enumitem}
\usepackage{bm}
\usepackage{eurosym}
\usepackage{graphicx}
\usepackage{epstopdf}
\usepackage{subfigure}
\graphicspath{{./FIGURES/}}        
\usepackage{psfrag}
\usepackage{booktabs,multirow,longtable}
\usepackage[titletoc,page]{appendix}
\usepackage{natbib}

\newcommand{\cmmnt}[1]{}

\hypersetup{
colorlinks=true,
citecolor=blue,
linkcolor=magenta
}

\newtheorem{theorem}{Theorem}[section]

\title{\bf Chebyshev Greeks\\ [1ex] \Large Smoothing Gamma without Bias}
\author{
Andrea Maran\thanks{Independent consultant. Email: {\tt andre.maran95@gmail.com}.}
\ \ \
Andrea Pallavicini\thanks{Department of Mathematics, Imperial College, London SW7 2AZ, UK and Intesa Sanpaolo, Largo Mattioli 3, 20121 Milano, Italy. Email: {\tt andrea.pallavicini@intesasanpaolo.com}.}
\ \ \
Stefano Scoleri\thanks{Be Management Consulting, Piazza Affari 2, 20123 Milano, Italy. Email: {\tt s.scoleri@be-tse.it}, corresponding author.}
}

\date{
\small First Version: June 7, 2021. This Version: \today
}

\begin{document}

\maketitle

\begin{abstract}

The computation of Greeks is a fundamental task for risk managing of financial instruments. The standard approach to their numerical evaluation is via finite differences. Most exotic derivatives are priced via Monte Carlo simulation: in these cases, it is hard to find a fast and accurate approximation of Greeks, mainly because of the need of a tradeoff between bias and variance. Recent improvements in Greeks computation, such as Adjoint Algorithmic Differentiation, are unfortunately uneffective on second order Greeks (such as Gamma), which are plagued by the most significant instabilities, so that a viable alternative to standard finite differences is still lacking. We apply Chebyshev interpolation techniques to the computation of spot Greeks, showing how to improve the stability of finite difference Greeks of arbitrary order, in a simple and general way. The increased performance of the proposed technique is analyzed for a number of real payoffs commonly traded by financial institutions.      

\end{abstract}

\bigskip
\bigskip
\bigskip
\bigskip

\noindent {\bf AMS classification codes:} 41A10, 68U20, 68W25, 90C59.

\newpage
\tableofcontents
\vfill

\section{Introduction}
\label{sec:introduction}
The evaluation of sensitivities of financial derivatives with respect to specific market and model parameters (so called Greeks) is a fundamental task both for Front Office and Risk Management departments of a financial institution.
For example, Greeks are used on a daily basis by derivatives traders to hedge their books against movements of the market. In addition to the well known problem of managing risks, regulators are also increasingly pointing towards a sensitivity-based representation of financial risks (see e.g. FRTB, SIMM, SA-CVA). All these reasons entail the need for a fast and accurate computation of Greeks.

On a mathematical standpoint, Greeks are basically derivatives of the pricing function with respect to given variables.
The standard approach to the numerical evaluation of derivatives is via finite differences (FD). This approach implies that the pricing function is called multiple times, on each bumped scenario: this could be particularly expensive, considering that complex payoffs are usually priced via Monte Carlo (MC) simulation. Moreover, when finite differences are coupled with Monte Carlo, the bias-variance problem can lead to relevant numerical instabilities, especially for second order Greeks \citep{Jac02,Gla03}: if one tries to decrease the size of the bump, in order to reduce the bias coming from the approximation of the derivative with a finite difference, then the variance of the Monte Carlo estimator of the derivative increases, making the result noisy. In practice, a tradeoff must be empirically found in the choice of the bump. As a result, computing Greeks in a fast and accurate way turns out to be a demanding task, possibly threatening the reliability of calculated hedge ratios. 

In the last decade, the introduction of Adjoint Algorithmic Differentiation (AAD) in the financial industry solved both the speed and accuracy problems for first order greeks \citep{GilGla06,Gri08,Cap11,Nau12,Sav18}: unbiased estimates of an arbitrary number of Greeks can be obtained at a cost which is comparable to the evaluation of the pricing function itself. More precisely, it can be proven that, given a scalar function of many variables, the computational cost of evaluating its gradient with AAD is approximately four times the cost of evaluating the function alone, independently of the number of derivatives to compute. Unfortunately, this result does not generalize to second order derivatives, unless in diffusive settings thanks to a link with first order Greeks \citep{Dal20}. Moreover, adjoint techniques may suffer from numerical instabilities, particularly for payoffs with discontinuities. All trivial tricks, such as increasing the bump or smoothing the discontinuity (for example, replacing indicator functions by tight call spreads), are able to reduce the MC noise but invariably add a bias to the result. Therefore, if an accurate gamma is needed, one is usually forced to use a huge number of MC paths, thus worsening the performance of the computation. At this regard, we also notice that usual techniques aimed at accelerating MC convergence, such as Quasi Monte Carlo with Sobol' sequences, are often uneffective on gamma (see \cite{BiaKuc15} for details). 

In this work, we discuss an application of Chebyshev interpolation to the computation of Greeks of arbitrary order, aiming to improve the performance of finite differences. Chebyshev interpolation techniques have recently gained interest in Finance and, in particular, in risk management, because of their ability to boost the performance of counterparty risk computations \citep{Gla16,Rui18,Gla19a,Gla19b,Gla20a,Gla20b,Rui20,Rui21a,Rui21b}. The underlying idea is to approximate, under suitable regularity conditions, the original pricing function $f(x)$ with a polynomial $p(x)$, interpolating the values of $f$ on a grid of $n$ points $\{x_i\}_{i=1,\ldots,n}$ in a given interval $[a, b]$ for the parameter $x$ to be varied. If the points $\{x_i\}$ are chosen to be the Chebyshev points and $f$ is analytical, the approximated function $p_n$ exponentially converges to the original one for increasing $n$. Remarkably, this is still true for the derivatives: therefore, in the case of Chebyshev interpolation, the derivatives of the polynomial interpolant, $p^{(m)}$, are also a good approximation to the actual derivatives $f^{(m)}$. We refer to section \ref{sec:chebyshev} for all details. Given a financial product described by a pricing function $f$, we propose to approximate its Greeks at the point $x_0$ (e.g. $x_0$ can be the spot price of the underlying asset, in the case of delta and gamma) with $p^{(m)}(x_0)$. These derivatives can be computed in an effective way thanks to the barycentric formula. In particular, we provide some heuristic rules to choose $n$ and $[a, b]$ so that the approximation error is minimized and possible singularities are correctly handled.     

The paper is organized as follows: in section \ref{sec:chebyshev} we introduce the theoretical framework and the proposed methodology, in section \ref{sec:numeric} we present some tests on real payoffs and assess the performance of the proposed methodology with respect to standard finite differences, while in section \ref{sec:conclusion} we conclude and suggest some directions of future work. Some technical considerations on the errors of polynomial interpolation techniques applied to MC prices and Greeks are provided in the appendices.

\section{Chebyshev Methods for Price and Greeks Approximation}
\label{sec:chebyshev}
This section is devoted to some theoretical considerations which are set at the ground of our proposal for an effective computation of Greeks. In section \ref{sec:fd} we briefly review standard finite difference techniques, with particular focus on their interaction with MC simulations. In section \ref{sec:chebgreeks} we recall some key results on polynomial interpolation. In section \ref{sec:adaptive} we introduce our methodology based on Chebyshev interpolation techniques for Greeks computation, together with some heuristics for obtaining stable results in general situations.

\subsection{Finite Differences}
\label{sec:fd}
Consider a function $f:U(x_0)\to\mathbb{R}$, defined on some neighbourhood $U(x_0)=[x_0 - a, x_0 + a]$ of $x_0$. Let $f$ be differentiable at least twice\footnote{It is easy to extend the discussion to higher derivatives, but for financial applications we are only interested in derivatives up to second order.} in $x_0$. In the vast majority of pricing applications, when finite differences are chosen to approximate Greeks, 3-point central differences are used as they provide second order approximations of derivatives in the bump $h$, with only two additional function evaluations (see e.g. \cite{Gla03}):
\begin{align}
\label{eq:3ptsFD1}
f'(x_0) =\,& \frac{1}{2h}\Big[-f(x_0-h) + f(x_0+h)\Big] + \mathcal{O}(h^2),\\
\label{eq:3ptsFD2}
f''(x_0) =\,& \frac{1}{h^2}\Big[f(x_0-h) -2f(x_0) + f(x_0+h)\Big] + \mathcal{O}(h^2).
\end{align} 
If the function $f$ is computed by a MC simulation, equations (\ref{eq:3ptsFD1}, \ref{eq:3ptsFD2}) can still be used, but the variance of the estimation of the derivatives increases when reducing $h$, which is in contrast with the need to choose a small value for $h$ to reduce the finite difference bias. In particular, if the same MC seed is used for all function evaluations, we have
\begin{equation}\label{eq:biasvar}
\text{Bias}[f', f''] = \mathcal{O}(h^2), \qquad \text{Var}[f'] = \mathcal{O}(h^{-1}), \qquad \text{Var}[f''] = \mathcal{O}(h^{-3}).
\end{equation}
Equation (\ref{eq:biasvar}) gives a clear hint on why second order derivatives are so hard to compute with FD in a MC approach. This is the well known bias-variance problem. One way to tackle this problem is to switch to $n$-point central differences, for $n>3$. In this case the bias is of order $\mathcal{O}(h^{n-1})$ while the variance is unaffected. For example, the 7-point differences are given by the following formulas:
\begin{align}
\label{eq:7ptsFD1}
f'(x_0) =\,& \frac{1}{60h}\Big[-f_{-3} + 9f_{-2} - 45f_{-1} + 45f_1 - 9f_{2} + f_{3}\Big] + \mathcal{O}(h^6),\\
\label{eq:7ptsFD2}
f''(x_0) =\,& \frac{1}{180h^2}\Big[2f_{-3} - 27f_{-2} + 270f_{-1} - 490f_0 + 270f_{1} - 27f_{2} + 2f_{3}\Big] + \mathcal{O}(h^6).
\end{align}
where $f_{\pm k}:=f(x_0\pm kh)$. Therefore, one can increase the bump size to decrease the variance without impacting too much on the bias. This comes at the cost of additional revaluations of the pricing function.

\subsection{Polynomial Interpolation}
\label{sec:chebgreeks}
In this section, we restrict ourselves to functions $f$ defined on $[-1,1]$. The general case can be easily obtained after an affine transformation and the following results will be unaffected. We refer to \cite{Tre18} for a complete introduction on polynomial interpolation methods.

The formulas for $n$-point central differences are usually derived by Taylor-expanding the function $f$ around $x_0$ up to order $n-1$. However, they can also be obtained computing the derivatives, at $x_0$, of the Lagrange polynomial interpolating points $f(x_k)$ on a uniform grid of $n$ points around $x_0$, with spacing $h$. 

The Lagrange interpolant is the unique polynomial $p_{n-1}$ of degree at most $n-1$ which satisfies $f(x_k)=p_{n-1}(x_k)$ for each $x_k$ in the interpolation grid $\{x_i\}_{i=0,\ldots,n-1}$. It is given by
\begin{equation}\label{eq:laginterp}
p_{n-1}(x) = \sum_{k=0}^{n-1} f(x_k)\,\ell_k(x)
\end{equation}
where $\ell_k(x) = \prod_{j\ne k}^{n-1} \frac{x - x_j}{x_k - x_j}$ are the Lagrange polynomials, i.e. the unique
polynomials of degree $n-1$ taking the value 1 at $x_k$ and 0 at the other points $x_i$ of the grid. We notice that formula (\ref{eq:laginterp}) is not limited to uniform grids, but holds for generic grids. 

The Lagrange interpolant can be evaluated effectively at any point $x\in[-1,1]\setminus \{x_i\}$ through the barycentric formula:
\begin{equation}\label{eq:barycentric}
p_{n-1}(x) = \frac{\sum_{k=0}^{n-1}\frac{w_k\,f(x_k)}{x-x_k}}{\sum_{k=0}^{n-1}\frac{w_k}{x-x_k}},\qquad w_k := \frac{1}{\prod_{j\ne k}^{n-1} (x_k-x_j)}.
\end{equation}
As pointed out in \cite{BerTre04}, the barycentric formula is very quick, since it evaluates the polynomial in $\mathcal{O}(n)$ flops after the barycentric weights $w_k$ have been pre-computed, and also very stable in many cases (including Chebyshev grids), since its scale-invariances avoid underflow or overflow in the computation of the weights.

The evaluation of derivatives of a Lagrange interpolant is very easy and doesn't require any additional evaluation of the function $f$: differentiating equation (\ref{eq:laginterp}) at grid points $\{x_i\}$ yields
\begin{equation}\label{eq:lagder}
p_{n-1}^{(m)}(x_i) = \sum_{k=0}^{n-1} f(x_k)\,\ell_k^{(m)}(x_i) = \sum_{k=0}^{n-1} D_{ik}^{(m)}\,f(x_k)
\end{equation}
which is simply the multiplication of a $n\times n$ differential matrix $D^{(m)}_{ik}:=\ell_k^{(m)}(x_i)$ with the vector containing the values of the function $f$ at the grid points. The differential matrices depend only on the grid points $\{x_i\}$ and are given by the following recursive formula (see \cite{Wel97}):
\begin{equation}\label{eq:diffmat}
D_{ik}^{(0)} =  \delta_{ik}, \qquad D_{ik}^{(m)} = 
\begin{cases}
\frac{m}{x_i-x_k}\,\left(\frac{w_k}{w_j}\,D_{ii}^{(m-1)} - D_{ik}^{(m-1)}\right) & \mbox{if } i\ne k\\
 & \\
-\sum_{j\ne i}^{n-1} D_{ij}^{(m)} & \mbox{if } i = k  
\end{cases}
\end{equation}
The value of $p_{n-1}^{(m)}(x)$ at a generic point can then be obtained via the barycentric formula (\ref{eq:barycentric}), replacing $f(x_k)$ with $p_{n-1}^{(m)}(x_k)$ as computed with (\ref{eq:lagder}).

It is known that, for generic grids including the uniform grid, polynomial interpolations have bad convergence properties (see e.g. the duscussion in \cite{Rui18}). On the contrary, lagrangian interpolation on the so called Chebyshev points (or other points properly clustered at the endpoints of the interval) enjoys optimal convergence properties, at least for some classes of functions. 

Let $\{z_i\}_{i=0,\ldots,n-1}$ be $n$ equispaced points on the upper unit circle in the complex plane. Chebyshev points are defined as their projections on the real line:
\begin{equation}\label{eq:chebpts}
x_k = \text{Re}[z_k] = \cos\left(\frac{k\,\pi}{n-1}\right),\qquad k=0,\ldots,n-1
\end{equation}
The following result shows that exponential convergence of the Chebyshev interpolant and all its derivatives is guaranteed for analytic functions (see \cite{Tre18}, chapter 21).
\begin{theorem}\label{theo:chebconv}
	Let $f$ be an analytic function on $[-1,1]$ which is analytically continuable to the closed Bernstein ellipse $\bar{E}_\rho$ of radius $\rho>1$. Then, for each integer $m>0$, there exists a constant $C>0$ such that
	\[||f^{(m)} - p_{n-1}^{(m)}||_\infty \le C\,\rho^{-n}\, .\]
\end{theorem}
\noindent Chebyshev interpolants can be expressed in the basis of Chebyshev polynomials $\{T_k(x)\}$, with coefficients $\{c_k\}$ given as Fast Fourier Transforms of $\{f(x_k)\}$. However, the best way to evaluate Chebyshev interpolants and their derivatives is via the barycentric formula, where the weights can be analytically evaluated as:
\begin{equation}\label{eq:chebweights}
w_k = \begin{cases} \frac{1}{2}\,(-1)^k & \mbox{if } k=0,\,n-1\\ (-1)^k & \mbox{otherwise} \end{cases}
\end{equation}

\subsection{Chebyshev Greeks with Adaptive Domains}
\label{sec:adaptive}
We are now ready to formulate our proposal for Greeks computation. Let $f(x)$ be the price of a financial product as a function of the parameter $x$. We want to numerically evaluate its derivatives at some point $x_0$. In the applications of the present work, $x_0$ will be the spot price of one of the underlying assets, $f'(x_0)$ will be the delta and $f''(x_0)$ will be the gamma.

We approximate the Greeks $f^{(m)}(x_0)$ with the derivatives $p^{(m)}(x_0)$ of a Chebyshev interpolant of $f(x)$ in some region around $x_0$. Among other polynomial interpolations, we pick Chebyshev because of its optimal properties described in section \ref{sec:chebgreeks}. The detailed steps of the methodology are as follows:
\begin{enumerate}
	\item Choose the interpolation domain $H = [x_0-a,\, x_0+a]$ and the number $n$ of Chebyshev points. Below we present some heuristics to find the optimal values of $a$ and $n$. It is convenient to choose $n$ as an odd integer, so that the point $x_0$ is included in the Chebyshev grid and the price $f(x_0)$ will be obtained while building the interpolator (see step 2) without additional evaluations.
	
	\item Build the Chebyshev interpolator. This amounts to:
	\begin{enumerate}[label=(\roman*)]
		\item compute Chebyshev points $\{x_k\}_{k=0,\ldots,n-1}$ via (\ref{eq:chebpts}) and map them from the unit interval to $H$ with the appropriate affine transformation;
		\item compute barycentric weights via (\ref{eq:chebweights});
		\item compute the differential matrices up to the desired order $m$ via (\ref{eq:diffmat});
		\item evaluate the original pricing function on the Chebyshev points to obtain the interpolation nodes $\{f(x_k)\}_{k=0,\ldots,n-1}$. This is usually the most expensive step. If $f$ is evaluated through MC simulation, we should fix the seed at each revaluation, so as not to add spurious discontinuities.
	\end{enumerate}

	\item Obtain the desired Greeks as $p^{(m)}(x_0)$ using (\ref{eq:lagder}) and, if $n$ is even, (\ref{eq:barycentric}) replacing $f(x_k)$ with $p^{(m)}(x_k)$. This step is almost instantaneous.  
\end{enumerate}	    
The above procedure depends on the choice of two adjustable parameters\footnote{We notice that also standard finite differences techniques depend on the choice of two paramters: the number $n$ of points, often set to 3, and the bump size $h$, which is also related to the size of the approximation domain.}: the number of nodes $n$ and the domain size $a$. The number of nodes $n$ should be greater than 3 (otherwise the method would degenerate to standard central differences) but not too high, in order to keep low the building time of the interpolator. We found empirically that $n=7$ is a good compromise in most situations.

Having fixed $n$, we are left with the parameter $a$ to be chosen. The choice of a small value for $n$ is justified when the conditions of theorem \ref{theo:chebconv}, basically analyticity of $f$ inside $H$, are satisfied: therefore, the choice of $a$ should be guided by the possible presence of singularities either in the price function or in its derivatives. Notice that, even when the underlying function $f$ is analytic, if it is estimated via MC simulation then the estimator is no longer analytic and we cannot strictly rely on theorem \ref{theo:chebconv}: however, the results presented in appendix \ref{sec:convergence} empirically support the idea that good convergence properties still hold in a MC setting. Apart from this fact, true singularities are usually present only at known fixing dates and are located at known levels (e.g. barriers, strikes, etc.). Measurable singularities can be removed with independent techniques. Even though actual singularities are formally present only at specific dates, interpolation may struggle also when the singularity date approaches, even though the pricing function is smooth. Indeed, around the level of an approaching singularity, the price and its derivatives show significant oscillations: the interpolation domain should, therefore, capture this behaviour, avoiding regions where the function is almost flat. One possible solution is to adjust the size of $H$ according to the ``time to next singularity'' $\tau$ and underlying volatility $\sigma$. This is motivated by the following argument: assuming that singularities are generated by digital features (indicators present in the payoff), for digital options in Black model the scale of the singularity is given by $\sigma\sqrt{\tau}$ (Black model is only used to estimate Chebyshev parameters, but the method works with general dynamics: in particular, in section \ref{sec:numeric} we consider models with local and stochastic volatility). As we move away from the singularity, we can use higher domain sizes, thus exploiting the good properties of Chebyshev interpolator to reduce the MC variance of standard finite differences. A possible implementation of this time- and state-adaptive strategy is the following:
\begin{enumerate}
	\item Let $\tau$ be the time to next singularity date $T$, $\{b_i\}_{i=1,\ldots,B}$ the positions of the singularities at time $T$, $\sigma$ the ATM volatility of the underlying asset, directly estimated from the market quotes of plain vanilla options, and $x_0$ its spot price.
	\item Define:
	\begin{enumerate}[label=(\roman*)]
		\item $a_\tau := \alpha\,x_0\,\sigma\,\sqrt{\tau}$, for some $\alpha\in[1, 2]$
		\item $d_i := |x_0 - b_i|$, $\forall i=1,\ldots,B$
		\item $a_b := \min_{i=1,\ldots,B} \frac{1}{2}(d_i - a_\tau)^+$
	\end{enumerate}
	\item Set the size of the interpolation domain $H$ equal to:
	\begin{equation}\label{eq:adaptivebumps}
	a = \min\Big(\max\left(a_b + a_\tau,\, a_{min}\right),\, a_{max}\Big)
	\end{equation}
	where $a_{min}$ and $a_{max}$ are appropriate bounds. For example, one can set $a_{min} = \left\lfloor \frac{n}{2}\right\rfloor\cdot h$, where $h$ is the characteristic bump of standard 3-point central differences, and $a_{max}$ large enough to span all relevant features of the payoff (strikes, barriers, etc.). 
\end{enumerate}

\section{Numerical Investigations}
\label{sec:numeric}
In this section, we perform some numerical experiments to assess the effectiveness of the Chebyshev methodology introduced in section \ref{sec:adaptive} and compare it to standard finite differences. We aim to show that, within MC simulations, more stable Greeks can be obtained at a reduced computational cost and without significant biases. Indeed, as explained in section \ref{sec:fd}, meaningful results for second order Greeks can be achieved only resorting to a huge number of MC paths, with standard techniques. On the contrary, with Chebyshev Greeks, while the number of re-pricings is slightly increased, the number of MC paths for each revaluation can be dramatically reduced while preserving accuracy.

We consider two types of exotic derivatives under complex pricing models: FX target redemption forwards (TARFs) under the Stochastic Local Volatility model by \cite{TatFis10} and Equity autocallables under a multi-asset Local Volatility model \citep{Dup94,DerKan94}. Both payoffs can show singularities due to the presence of different types of barriers, so we can test the performance of the adaptive method outlined in section \ref{sec:adaptive}. Additionally, in appendix \ref{sec:errors} we consider a digital option under the Black model, as a textbook example which allows to perform a better error analysis.

\subsection{Target Redemption Forwards}
\label{sec:TARF}
We consider a TARF on EUR/USD exchange rate $S$, with weekly put-like coupons with strike $K=1.15$, which are paid until a maximum payout $\theta$ (target) is reached. Negative coupons payments are triggered by a Knock-In barrier set at $B_{KI}=1.19$. Additionally, a Knock-Out barrier at $B_{KO}=1.135$ is present. At each coupon fixing date $T_i$, the TARF payoff can be written as:
\begin{equation}
\Pi(T_i) = \frac{K-S_{T_i}}{K\,S_{T_i}}\,\Big(\mathds{1}_{\{S_{T_i}\le K\}} + \mathds{1}_{\{S_{T_i} > B_{KI}\}}\Big)\,\mathds{1}\Big\{\sum_{j=1}^i(K-S_{T_j})^+ < \theta\Big\}\,\Big(1 - \min\Big\{1,\, \sum_{j=1}^i\mathds{1}_{\{S_{T_j} < B_{KO}\}}\Big\}\Big)\,N_i
\end{equation}
where $N_i$ are coupon notionals. There are 70 remaining coupons and the residual target is 0.2.
 
\begin{figure}
	\centering
	\subfigure[Delta]{\includegraphics[width=7in,height=3.5in,keepaspectratio=true]{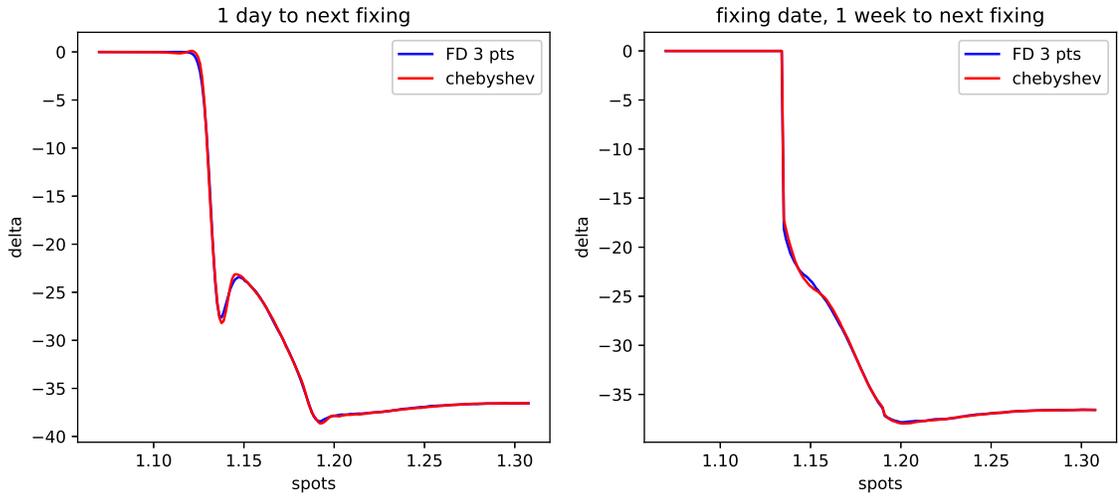}}\\
	\subfigure[Gamma]{\includegraphics[width=7in,height=3.5in,keepaspectratio=true]{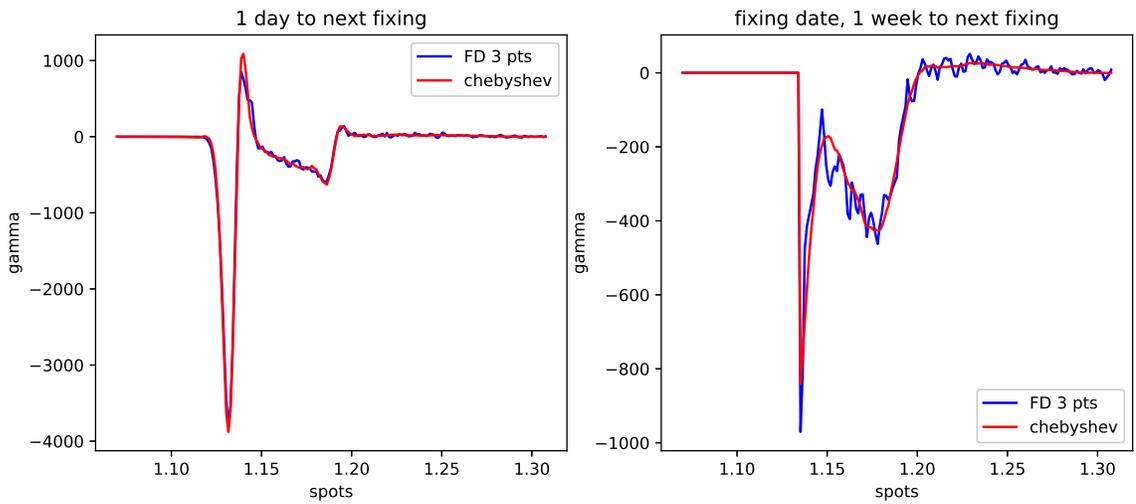}}
	\caption{Delta (a) and Gamma (b) of a TARF for 200 spot levels. Comparison between 3-point central differences with $10^6$ MC paths (blue) versus 7-points Chebyshev Greeks with $3\cdot 10^5$ MC paths and with time- and state-adaptive interpolation domain. Singularities show up at 1.135 and 1.19 spot levels. Results shown for 1-day (left) and 1-week (right) to a singularity date.} \label{Fig:TARF}
\end{figure}

In Figure \ref{Fig:TARF} we show the results of delta and gamma for different spot levels and evaluation dates. The details of the computation are summarized in Table \ref{Tab:results}. The improved stability of Chebyshev gamma is evident. We highlight that FD Greeks were obtained with 1M simulation paths for each call to the pricing function, while Chebyshev Greeks with 300K paths: since 7 Chebyshev points were employed, against 3 points for central FD, overall we also reduced the computational time by one third. Given the superior accuracy of Chebyshev result, there is still room for further computational time savings, depending on the desired accuracy threshold. 

In order to assess the consinstency of our methodology, we proceed as follows. We check how the computed delta (resp. gamma) is good at explaining the actual change in price (resp. delta), with the different methods. To this purpose, we define the following ``explanation errors'':
\begin{equation}\label{eq:explerr}
\varepsilon_M(\Delta) = \max_p \left|\Delta_M(S_p)\cdot dS_p - dP(S_p)\right|,\qquad \varepsilon_M(\Gamma) = \max_p \left|\Gamma_M(S_p)\cdot dS_p - d\Delta_M(S_p)\right|
\end{equation}  
where the spots $S_p$ run over the grid of points used for the tests, $P$, $\Delta_M$, $\Gamma_M$ are price, delta and gamma evaluated with the numerical method $M$. Notice that these explanation errors say nothing about numerical errors built in the computations (the latter are discussed in appendix \ref{sec:errors} for a simpler test case). They are simply used to check the self-coherence of Chebyshev Greeks: this is worth to be done, since at each spot $S_p$ a different interpolator is built.

We show the results in Table \ref{Tab:results}: the explanation  errors are comparable for all the analyzed methods.
\subsection{Autocallables}
\label{sec:AUTOC}
We now consider an autocallable option with memory, on a basket of two stocks: TELECOM and VODAFONE. The option pays a stream of coupons at times $T_i$, provided that the performance of the basket, with respect to a past strike date, is above $B_{coup}=90\%$. Additionally, an early-termination feature is present which gets activated if the basket performance is over $B_{call}=100\%$ at some $T_i$. Finally, there's a ``capital guarantee'' barrier on the last fixing date $T_N$ at $B_{gar}=60\%$. The basket performance is of ``worst-of'' type. At each coupon fixing date, the autocallable payoff can be written as:
\begin{equation}
\Pi(T_i) = \mathds{1}_{\{\tau > T_i\}} \left(\Bigg[N_i + \sum_{j=1}^{i-1}\Big(N_j - \Pi(T_j)\Big)\Bigg]\mathds{1}_{\{P(T_i)\ge B_{coup}\}} +\delta_{iN}\left(P(T_N)-1\right)\mathds{1}_{\{P(T_N)<B_{gar}\}} \right) + \mathds{1}_{\{\tau = T_i\}} R
\end{equation}
where $N_i$ are coupon notionals, $R$ is a rebate, $P(t) = \min\Big\{\frac{S_t^{tel}}{S_{t_ref}^{tel}}, \frac{S_t^{vod}}{S_{t_ref}^{vod}}\Big\}$ is the basket performance and $\tau = \min\{T_i:P(T_i)\ge B_{call}\}$ is the early-termination time. There are 7 remaining coupons, every 3 months. The initial fixings of the underlying assets, for the computation of the performances, are $S_{t_ref}^{tel}=0.48$ EUR for TELECOM and $S_{t_ref}^{vod}=1.3$ GBP for VODAFONE. The current spot price of VODAFONE is kept fixed to $S_0^{vod}=1.35$ GBP.

\begin{figure}
	\centering
	\subfigure[Delta TELECOM]{\includegraphics[width=7in,height=3.5in,keepaspectratio=true]{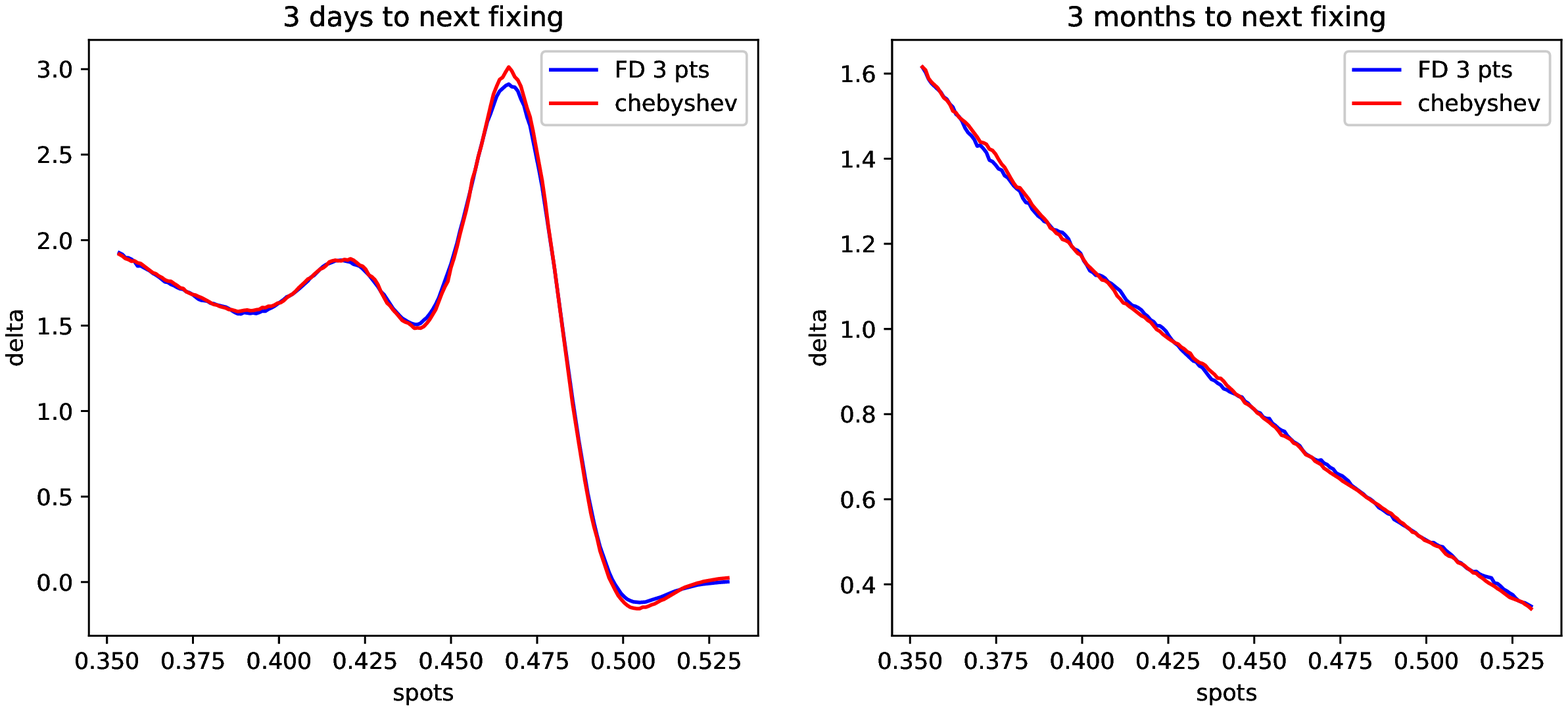}}\\
	\subfigure[Gamma TELECOM]{\includegraphics[width=7in,height=3.5in,keepaspectratio=true]{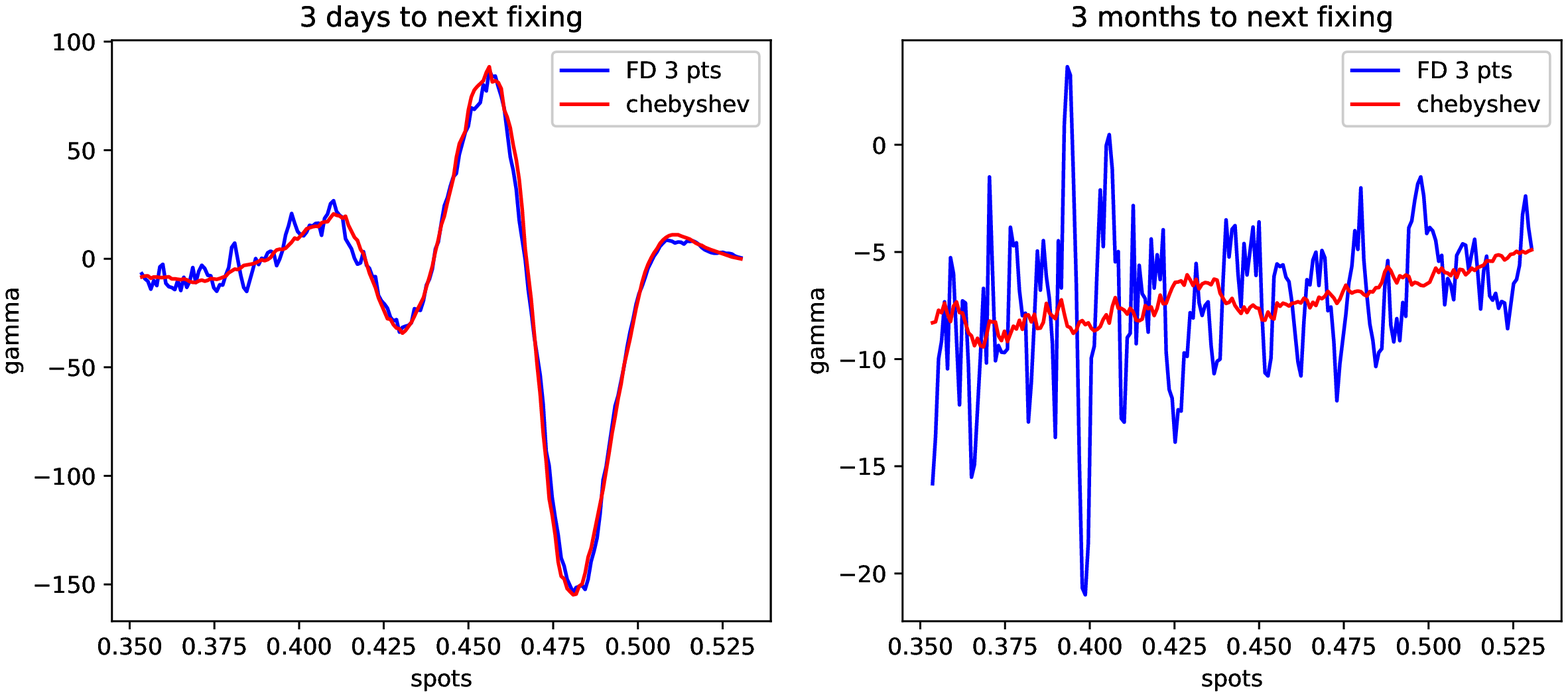}}
	\caption{Delta (a) and Gamma (b) of a worst-of autocallable on TELECOM and VODAFONE, for 200 spot levels of TELECOM. Comparison between 3-point central differences with $10^6$ MC paths (blue) versus 7-points Chebyshev Greeks with $3\cdot 10^5$ MC paths and with time- and state-adaptive interpolation domain. Singularities show up at 0.432 EUR and 0.48 EUR spot levels. Results shown for 3-day (left) and 3-month (right) to a singularity date.} \label{Fig:AUTOC}
\end{figure}
 
In Figure \ref{Fig:AUTOC} we show the results of delta and gamma for different evaluation dates and spot levels of TELECOM. The details of the computation are summarized in Table \ref{Tab:results}. Again, 300K MC paths were used with Chebyshev Greeks, aginst 1M paths with finite differences. The same comments on TARF results apply here and confirm the effectiveness of Chebyshev method.



\begin{table}
	\centering
	\begin{tabular}{c c c c c c c}
		\toprule
		\textbf{Payoff} & \textbf{MC paths} & \textbf{Greeks method} & \textbf{\# nodes} & \textbf{Bump min} & \textbf{Bump max} & \textbf{Expl. Err.}\\
		\midrule
		TARF & 1,000,000 & finite differences & 3 & 0.0025 & 0.0025 & 18.1\\
		TARF & 300,000 & adaptive Chebyshev & 7 & 0.0075 & 0.05 & 17.1\\
		AUTOC & 1,000,000 & finite differences & 3 & 0.01 & 0.01 & 0.02\\
		AUTOC & 300,000 & adaptive Chebyshev & 7 & 0.03 & 0.1 & 0.03\\
		\bottomrule
	\end{tabular}%
	\caption{Numerical details of the computations described in section \ref{sec:numeric}. ``Adaptive'' refers to the method described in section \ref{sec:adaptive} for the choice of the interpolation domain. ``Bump min'' and ``Bump max'' are expressed in terms of percentage of the spot: for FD greeks they coincide and are equal to $h/x_0$, while in the adaptive cases they are $a_{min}/x_0$ and $a_{max}/x_0$ respectively, see equation (\ref{eq:adaptivebumps}). Notations as in section \ref{sec:chebyshev}. ``Expl. Err.'' refer to $\varepsilon(\Gamma)$, as defined in (\ref{eq:explerr}). The worst case on all evaluation dates is shown. The errors were computed over a grid of 200 points. Delta explanation errors with different methods are indistinguishable.}\label{Tab:results}%
\end{table}%

\section{Conclusions and Further Developments}
\label{sec:conclusion}
In this work we presented a simple method to numerically evaluate delta and gamma Greeks of arbitrarily complex payoffs. It is based on Chebyshev interpolation over a suitable domain around the spot price. The degrees of freedom available in the choice of the interpolation domain enable us to use a low number of interpolation nodes, where the original pricer must be called. In order to do that, it is essential to adapt the size of the interpolation domain to the positions, in space and time, of possible singularities in the price or its derivatives. Considering some particularly exotic test cases, we showed that our methodology is able to substantially reduce the computational burden of standard techniques, based on finite differences, for gamma, while at the same time improving its numerical stability in MC simulations. This is due to the optimal convergence properties of Chebyshev interpolation.

The theory presented in this work is limited to the one-dimensional case. However, Chebyshev techniques (including barycentric interpolation and differential matrices) can be easily extended to $d$ dimensions with the use of Chebyshev tensors: appropriate tensor compression algorithms should be used to handle the high-dimensional cases, see \cite{Gla19b}. This is not striclty necessary in the applications presented here: after all, Greeks are partial derivatives with respect to single parameters. Nevertheless, the multi-dimensional extension is intersting to be explored and would offer the possibility to effectively compute cross-gammas, beside gammas.

\section*{Acknowledgments}
We thank Ignacio Ruiz and Mariano Zeron for fruitful discussions on the topics regarding Chebyshev interpolation covered in this paper. We also thank Giulio Sartorelli and Riccardo Longoni for useful collaborations in the development of this research. 

\section*{Disclaimer}
The views expressed here are those of the authors and do not represent the opinions of their employers. They are not responsible for any use that may be made of these contents.

\appendix
\section{Convergence of Polynomial Interpolants with MC Errors}
\label{sec:convergence}
Theorem \ref{theo:chebconv} is not strictly applicable in the case of pricing functions $f$ evaluated with a MC simulation. Indeed, the actual expected value of the discounted payoff is replaced by a sum over $N$ simulated paths and the estimated function, say $\bar{f}$, is no longer analytical because of the presence of MC errors. 

\begin{figure}
	\centering
	\subfigure[Analytical]{\includegraphics[width=3.2in,height=2.5in,keepaspectratio=true]{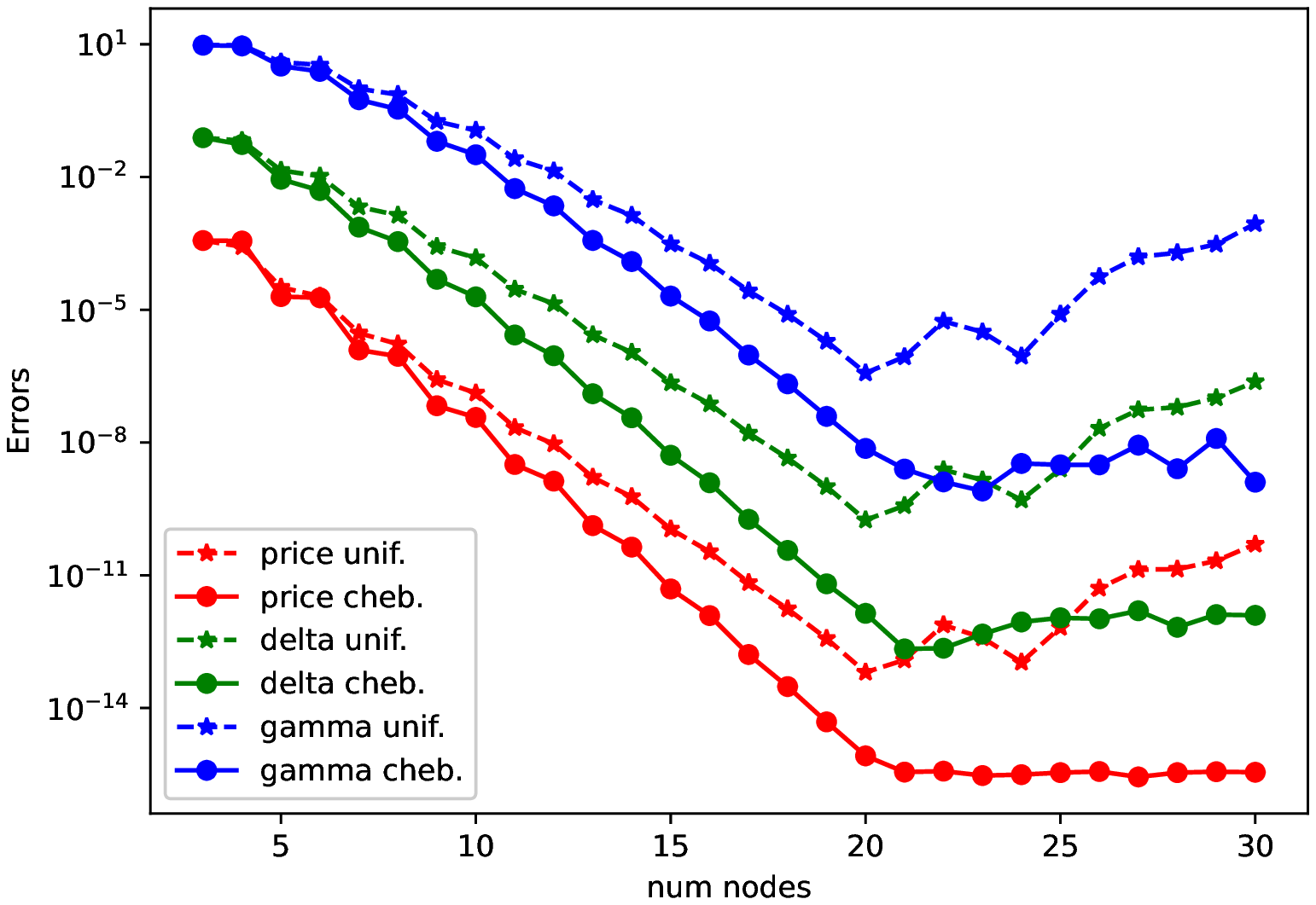}} 
	\subfigure[Monte Carlo]{\includegraphics[width=3.2in,height=2.5in,keepaspectratio=true]{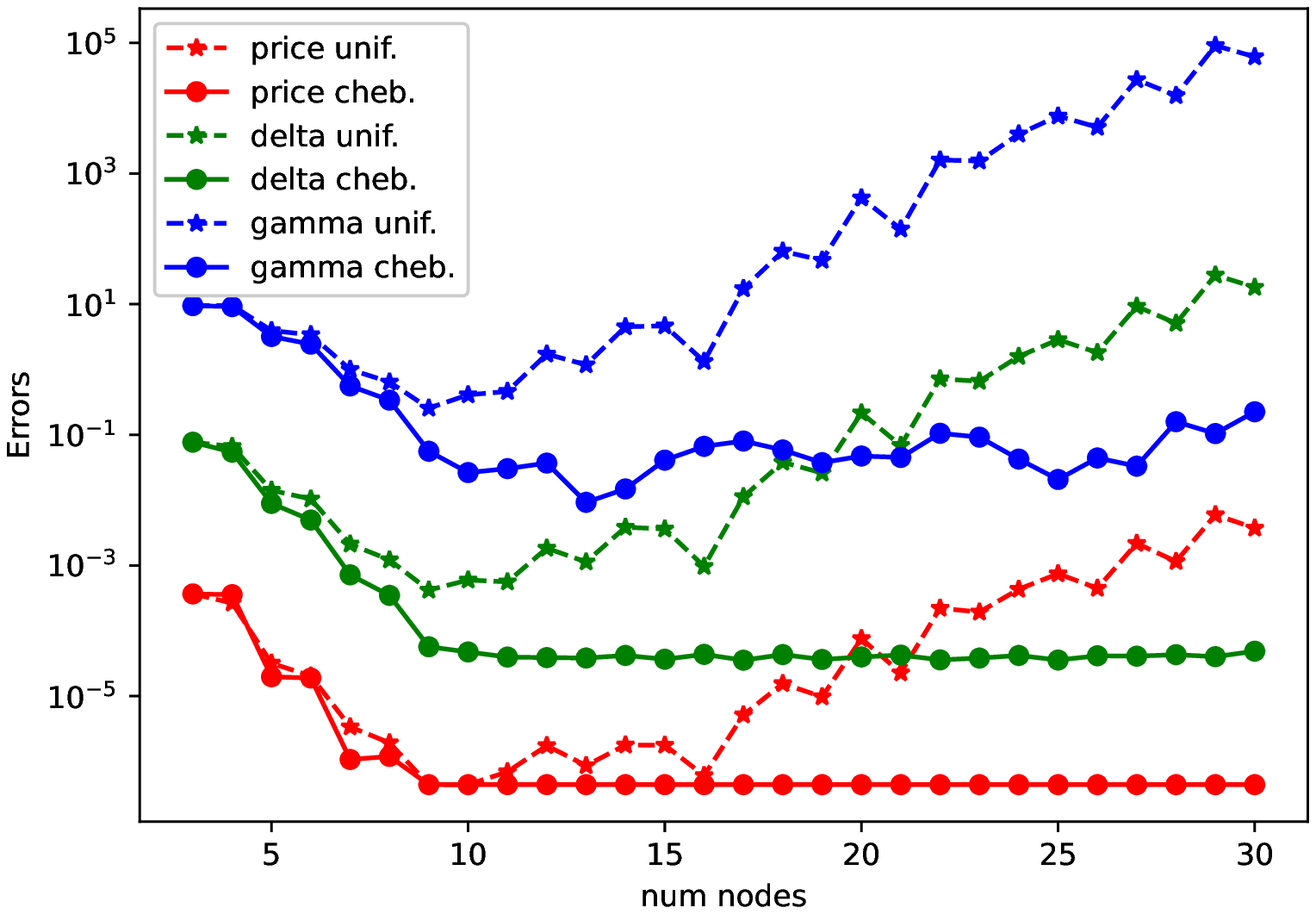}} 
	\caption{$L^{\infty}$ errors vs number of interpolation nodes for polynomial (uniform and Chebyshev) interpolants and their derivatives. The true values are given by exact Black formulas. Errors are evaluated over a grid of 1000 points around the strike. The polynomial interpolators were built using both the analytical pricer (a) and the MC pricer (b). The maximum MC error on the prices computed at nodal points is $2\cdot 10^{-6}$. Results shown for a call with strike $K=1$, time to maturity $T=0.1$, volatility $\sigma = 0.07$, risk-free rate $r=0$ and interpolation domain $H=[0.94, 1.01]$} \label{Fig:conv}
\end{figure}

Here, we consider a call option under Black model and empirically study the convergence of polynomial interpolants, say $\bar{p}^{(m)}_{n}$, of price $\bar{f}$ and Greeks $\bar{f}^{(m)}$ to the true values $f$ and $f^{(m)}$ as the number on interpolation nodes $n$ increases. As a measure of approximation error, we consider the $L^{\infty}$ distance between $\bar{p}_{n}^{(m)}$ and $f^{(m)}$, the latter being given by analytical Black formulas (see e.g. \cite{Wil06}), for $m=0,1,2$. We compare uniform and Chebyshev interpolators, over a domain with a size comparable to that provided by equation (\ref{eq:adaptivebumps}). The results shown in Figure \ref{Fig:conv} imply that Chebyshev interpolants exponentially converge up to the MC error, then they remain quite stable. On the contrary, uniform interpolators, after an initial convergence, diverge from true values because of the Runge phenomenon (see e.g. \cite{Tre18}, chapter 13).

\section{Error Analysis for Polynomial Approximation of Greeks}
\label{sec:errors}
Let us consider a digital option under Black model, whose payoff is defined as
\begin{equation}
\Pi(T) = \mathds{1}_{\{S_T > K\}}
\end{equation}
and repeat the same tests as those described in section \ref{sec:numeric}. Since analytical results $\Delta_{BS}$, $\Gamma_{BS}$ are available for Greeks in this simple case (see e.g. \cite{Wil06}), we can define the errors of our numerical approximations as follows:
\begin{equation}\label{eq:numerr}
\varepsilon_{n,\Delta}(S) = \left|p'_{n-1}(S) - \Delta_{BS}(S)\right|\, , \qquad \varepsilon_{n,\Gamma}(S) = \left|p''_{n-1}(S) - \Gamma_{BS}(S)\right|
\end{equation}

\begin{table}
\centering
\begin{tabular}{cccccccccc}
	\toprule
	Method & \# nodes &    Bump & Size &  Avg $\varepsilon_{\Delta}$ &  Std $\varepsilon_{\Delta}$ &  Max $\varepsilon_{\Delta}$ &  Avg $\varepsilon_{\Gamma}$ &  Std $\varepsilon_{\Gamma}$ &  Max $\varepsilon_{\Gamma}$ \\
	\midrule
	FD &       3 &  0.25\% &     - &         0.04 &         0.05 &         0.3 &        30.3 &        39.8 &       275.4 \\
	FD &       3 &  1\% &     - &         0.17 &         0.17 &         0.65 &         6.6 &         8.6 &        43.9 \\
	FD &       7 &  1\% &     - &         0.03 &         0.03 &         0.17 &         5.19 &         6.75 &        40.3 \\
	Cheb. &       7 & - &     3.32\% &         0.03 &         0.04 &         0.18 &         3.03 &         3.93 &        20.6 \\
	\bottomrule
\end{tabular}\caption{Error analysis for a digital call with strike $K=1$, time to maturity $T=0.1$, volatility $\sigma=0.07$ and risk-free rate $r=0$ under Black model. Averages, standard deviations and maxima of errors (\ref{eq:numerr}) over 2000 spot levels around the strike are shown. The finite difference bump $h$ and the Chebyshev domain size $a$ are given as percentages of the spot. Notations as in section \ref{sec:chebyshev}. 300K MC paths were used.}\label{Tab:err}
\end{table}

\begin{figure}
	\centering
	\subfigure[Delta]{\includegraphics[width=7in,height=4in,keepaspectratio=true]{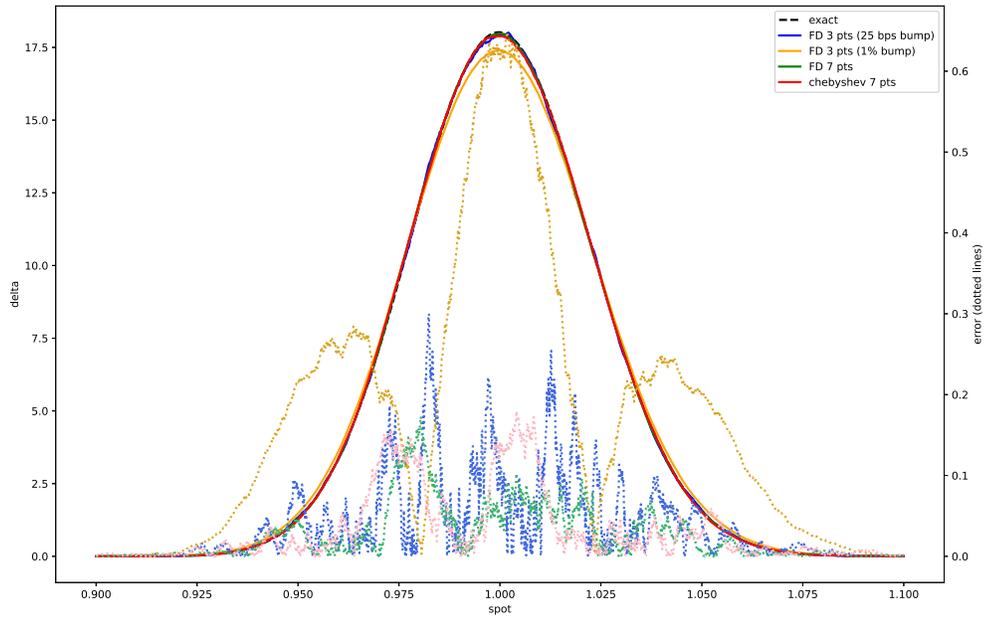}}
	\subfigure[Gamma]{\includegraphics[width=7in,height=4in,keepaspectratio=true]{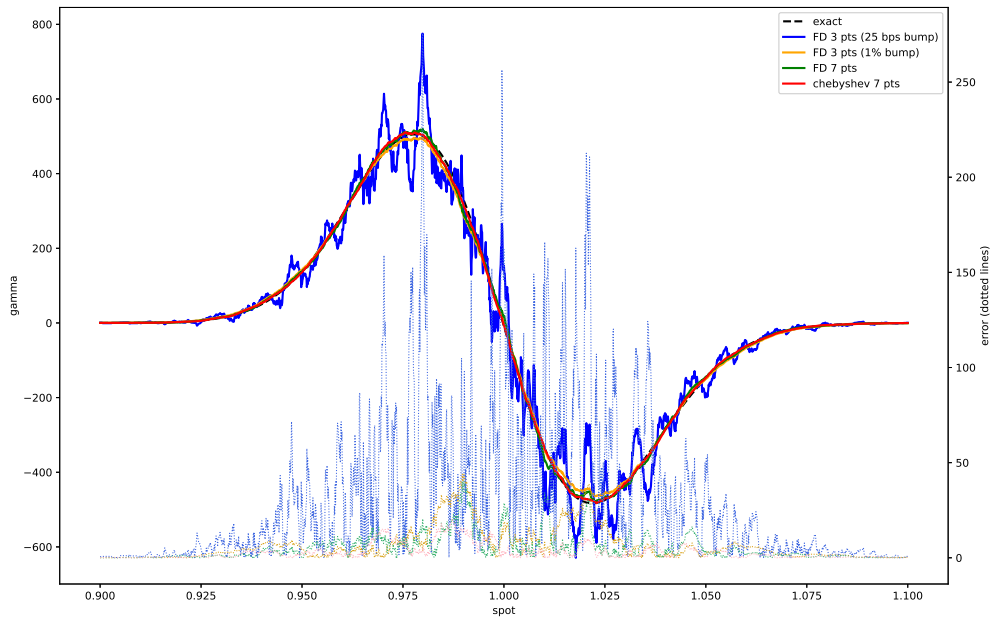}}
	\caption{Delta (a) and Gamma (b) values (left) and errors (right) of the digital call described in Table \ref{Tab:err} for 2000 spot levels around the strike. Various FD schemes are compared with the adaptive Chebyshev Greeks.} \label{Fig:err}
\end{figure} 

We aim to quantitatively measure the stability of the Chebyshev Greeks introduced in section \ref{sec:adaptive} with respect to finite differences. To this end, we compute errors (\ref{eq:numerr}) for different levels of the spot $S$ and provide some statistics. Results are summarized in Figure \ref{Fig:err} and Table \ref{Tab:err}. It turns out that 3-point finite differences display high variance or bias, depending on the choice of the bump. Both 7-point finite differences and Chebyshev significantly reduce the bias. The highest variance reduction is obtained with the adaptive Chebyshev method, especially for gamma, as it is evident from the standard deviation and the maximum of the errors.

\bibliographystyle{nonumber}

\end{document}